\documentclass[a4paper, 10pt, conference]{ieeeconf}      

\IEEEoverridecommandlockouts                              

\usepackage{graphicx}
\usepackage{siunitx}

\title{\LARGE \bf
Effect of Transducer Positioning in Active Noise Control
}

\author{Sajil C. K., Biji C. L and Achuthsankar S. Nair
\thanks{This work was supported by University Grants Commission(UGC), New Delhi, India}
\thanks{Authors are with Department of Computational Biology and Bioinformatics, University of Kerala, Kerala, India
        {\tt\small sajilckdcb@keralauniversity.ac.in}}
}

\begin{document}

\maketitle
\thispagestyle{empty}
\pagestyle{empty}

\begin{abstract}

Research in traditional Active Noise Control(ANC) often abstracts acoustic channels with band-limited filter coefficients. This is a limitation in exploring structural and positional aspects of ANC. As a solution to this, we propose the use of room acoustic models in ANC research. As a use case, we demonstrate anti-noise source position optimization using room acoustics models in achieving better noise control. Using numerical simulations, we show that level of cancellation can be improved up to 7.34 dB. All the codes and results are available in the Github repository https://github.com/cksajil/ancram in the spirit of reproducible research.

\end{abstract}

\section{INTRODUCTION}

Studies have proven that exposure to high sound pressure levels for extended periods of time can cause temporary or permanent hearing loss \cite{sierra2008acoustic}⁠. In the medical field, MRI scanners generate noise over 120 dB \cite{ravicz2000acoustic}⁠, which is often causing discomfort to the patients \cite{kannan2011efficient} as well as to doctors. There have been studies proving that the noise generated by fMRI scanners affect the scanning results also \cite{skouras2013fmri}. One solution to this is acoustic quiet zones generated by Active Noise Control(ANC), which have broad applications in medical, entertainment, automobiles \cite{samarasinghe2016recent}, airplanes, recording studios etc \cite{kuo1999active}. Recently, noise cancellation concepts were found application in synthetic biology also \cite{zechner2016molecular}.

\begin{figure}[!t]
\centering
\includegraphics[width=0.45\textwidth]{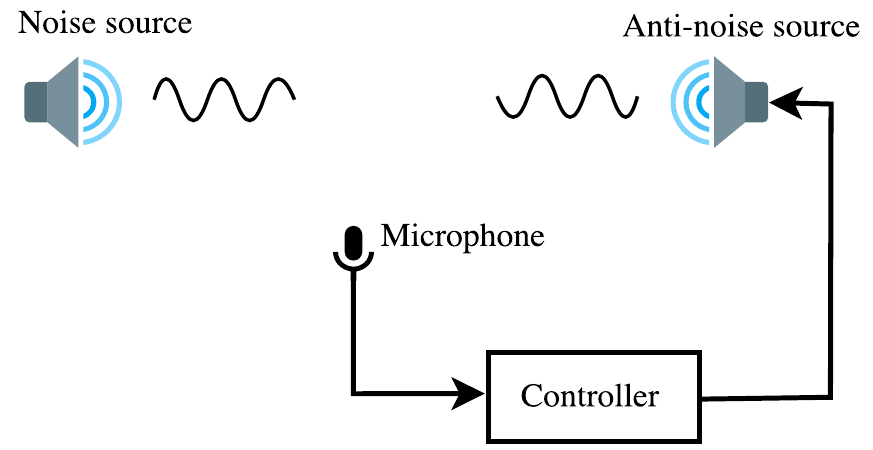}
\caption{Illustration of active noise control. The microphone receives the signal from the noise source and feeds into a controller. Controller analyses and drives the anti-noise source with phase inverted signal. Both signals cancel each other at the region of interest.}
\label{fig:ANCConcept}
\end{figure}

Passive noise control uses bulky sound absorbent materials to mask unwanted sound, whose performance degrades for low-frequency sounds \cite{kuo1999active}. ANC systems use the principle of canceling the sound wave with its inverse or anti-noise. The idea made its first appearance in the patent of Paul Lueg \cite{paul1936process}. Figure \ref{fig:ANCConcept} shows a graphical description of a simple active noise control system. Compared to passive noise control techniques, ANC techniques are well able to control low as well as high-frequency noise components. Also, it does not require bulky materials, making it portable.

In the existing ANC literature, which has a history of 7 decades \cite{george2013advances}, the acoustic channels are mostly modeled using Finite Impulse Response(FIR) filters \cite{veeravasantarao2008adaptive}. This approximation, used for ease of computation, limits researchers in studying the effects of physical aspects like transducer positioning, orientation, room reverberation, wall reflections, room structure etc. Positioning of transducers is an important parameter which is to be addressed in ANC \cite{kajikawa2012recent}. Room Impulse Response(RIR) is analogous to an Impulse Response(IR) in Linear Time invariant(LTI) systems. The IR is considered to characterize the behavior of an LTI system completely. With room acoustics models, it is easy to calculate RIRs between any two points. Though some work make use of measured RIRs for simulation and experimental validation \cite{reddy2011hybrid}\cite{ardekani2016statistical}, a large scale utilization of RIRs using room acoustic models in ANC is still an unexplored area. In the subject area of room acoustics, simultaneous multi channel RIRs were used to reconstruct room geometry \cite{dokmanic2013acoustic}. Inspired by their work, we focus to explore the use of RIRs in the Active Noise Cancellation(ANC) context. Our major contributions are:

\begin{itemize}

  \item The use of room acoustic models in ANC.
  
  \item A new workflow model for ANC research.

  \item Optimization of anti-noise source location which improves cancellation up to 7.34 dB.

\end{itemize}

Section \ref{ANC Preliminaries} introduces the fundamental of ANC concepts. Section \ref{Room Acoustics Models} gives a brief overview of room acoustics models. The simulation details can be found in Section \ref{Numerical Simulation Setup}. Results are discussed in detail in Section \ref{Results and Discussions}, including limitations of our approach and future scope. A tutorial style treatment of the research work is available at https://cksajil.github.io/ancram/ aiming at a novice reader.

\section{ANC Preliminaries}\label{ANC Preliminaries}

Existing ANC systems use a controller to predict the noise of interest which is to be canceled. The ANC system picks up the noise using single or multiple microphones and passes on to the controller. The acoustic channel from the noise source to the zone of interest is represented using a primary propagation path $P(z)$. Similarly, the acoustic channel from anti-noise source to the zone of interest is represented using a secondary path $S(z)$. The adaptive filter keeps on updating its coefficients so that the error signal at the zone of interest is minimized. The weights of the controller filter are updated using adaptive filter algorithms like Least Mean Square(LMS), Recursive Least-Squares(RLS) or Filtered-x Least Mean Square (FxLMS).

\begin{figure}[!t]
\centering
\includegraphics[width=0.45\textwidth]{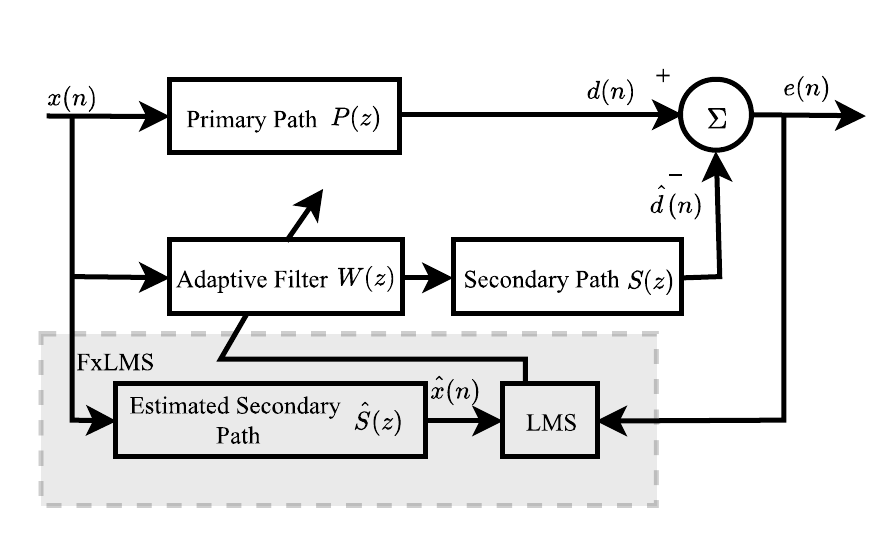}
\caption{Block diagram of FxLMS ANC System. $P(z)$ represent the acoustic channel from the noise source to the microphone. Similarly $S(z)$ represents the acoustic channel from anti-noise source to the microphone. The adaptive filter is controlled by the LMS algorithm which also considers the estimate of secondary path $\hat{S}(z)$ forming FxLMS.}
\label{fig:FxLMSBlock}
\end{figure}

Among the several types of ANC algorithms available, we tried our hypothesis with the popular and widely used FxLMS algorithm \cite{morgan1980analysis} due to its robustness and ease of computation. Figure \ref{fig:FxLMSBlock} shows block diagram of a traditional FxLMS-ANC system. The transformation happening to source signal $x(n)$, after passing through the primary propagation path $P(z)$, is modeled using the following equation.

\begin{equation}
d(n) = x(n)*p(n)
\end{equation}

The '$*$' here represents convolution operation. FxLMS algorithm also uses an estimate of the secondary path $\hat{S}(z)$, through which a filtered version of the source signal is calculated and passed on to the LMS filter.

\begin{equation}
\hat{x}(n) = \hat{s}(n)*x(n)
\end{equation}

The error signal $e(n)$ and the filtered reference signal $\hat{x}(n)$ are used to update the adaptive filter weights.

\begin{equation}\label{FxLMSEqn} w(n+1) = w(n)+\mu e(n)\hat{x}(n)
\end{equation}

Here, $w(n) = [w_{0}(n)\, w_{1}(n)\,...\,w_{L-1}(n)]$ are the filter weights, $L$ is the filter length and $\mu$ is the step size. In an ideal scenario, after several iteration, the predicted signal $\hat{d}(n)$ reaches inversely equal to $d(n)$ and $e(n)$ becomes zero. At this stage the filter weights $w(n)$ will be at its optimum converged range.

\section{Room Acoustics Models}\label{Room Acoustics Models}

Scientists working in acoustics or audio signal processing will often require a stage to test out and reproduce algorithms in specific acoustic scenarios. For this purpose, Room Acoustics Models(RAM) are often used, to recreate and reproduce room acoustics without any cross-model mismatch. Room acoustic models find its applications in beamforming \cite{dokmanic2015raking}, acoustic auralisation, speech and audio processing \cite{de2015modeling}, psychoacoustics studies etc \cite{wabnitz2010room}. 

\begin{figure}[!t]
\centering
\includegraphics[width=0.45\textwidth]{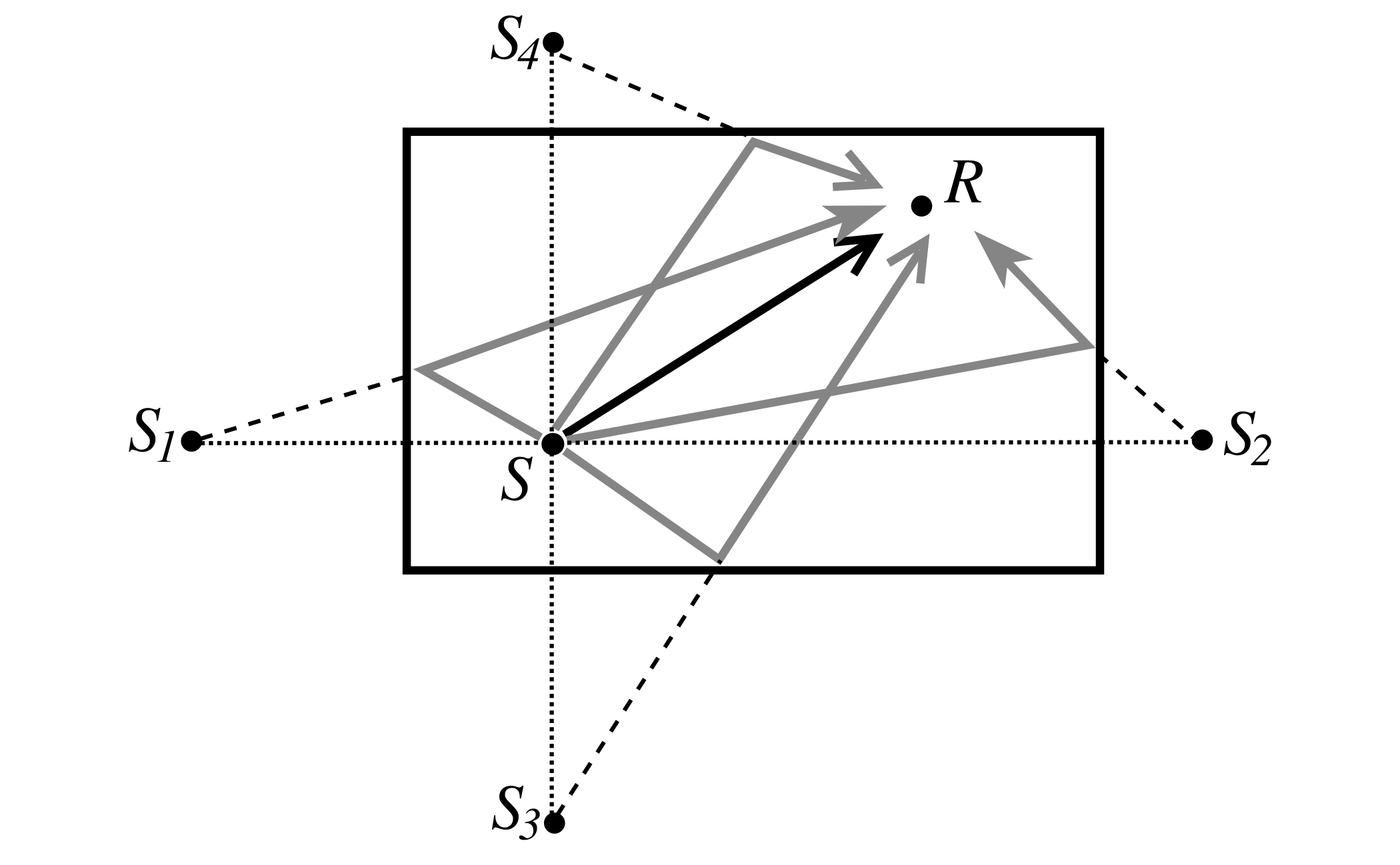}
\caption{Room impulse response as a superposition of direct sound and reflections. The direct sound (source $S$ to receiver $R$) is shown with solid arrow and reflections with gray arrows. The reflections of the direct sound appear to the receiver as coming from imaginary sources $S_1, S_2, S_3$ and $S_4$ which are scaled and delayed accordingly.}
\label{fig:RIRModel}
\end{figure}

A RIR model channel between a fixed source and a fixed receiver with direct path and reflections \cite{dokmanic2013acoustic}⁠. In RIR, apart from the direct sound, the receiver receives signals from different reflections. The acoustic energy gets attenuated after every reflection. A receiver at any point will be receiving direct sound superimposed with its reflections as reverberated sound.
Image method \cite{allen1979image} has been used widely to calculate RIRs for various room conditions.  In image source model, the reflections are considered coming from virtual sources. Figure \ref{fig:RIRModel} shows a diagrammatic representation of RIR for an arbitrary 2D room.

\section{Numerical Simulation Setup}\label{Numerical Simulation Setup}

\begin{figure}[!t]
\centering
\includegraphics[width=0.45\textwidth]{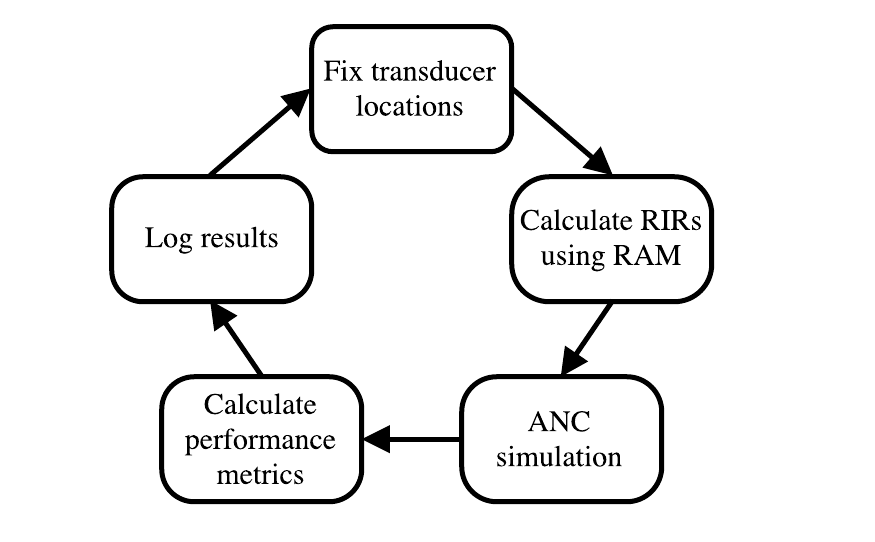}
\caption{Sequential steps in carrying out the simulation. For each anti-noise source location, RIRs were calculated using room acoustics models and passed onto ANC algorithm. The final canceled signal is analyzed and performance metrics(estimated attenuation and frequency specific attenuation) are logged for later reference.}
\label{fig:FlowDiagram}
\end{figure}

Figure \ref{fig:FlowDiagram} shows the flow diagram of the simulation. The room size was chosen arbitrarily to a size of $6\times4\times3$ meters. The z level of microphone, noise source and anti-noise source were fixed to a height of 1.53 for ease of computation. The x coordinate for noise source and microphone were chosen randomly from a uniform distribution within half-open interval $[1, 6)$. Similarly the y coordinate value for microphone and noise source were chosen randomly from a uniform distribution within half-open interval $[1, 4)$. The antinoise source positions were evenly spaced at a distance of \SI{11}{\cm}. For each position, anti-noise source is placed in that location and the primary path and secondary path, which are acoustic channels from the noise source to error microphone and anti-noise source to error microphone are calculated. The toolbox provided by Habets \cite{habets2006room} have been used for generating the RIRs. These RIRs were passed onto the ANC simulation program. A total of 100 independent Monte-Carlo simulations were carried out for different noise source and microphone positions.

\begin{figure}[!t]
\centering
\includegraphics[width=0.45\textwidth]{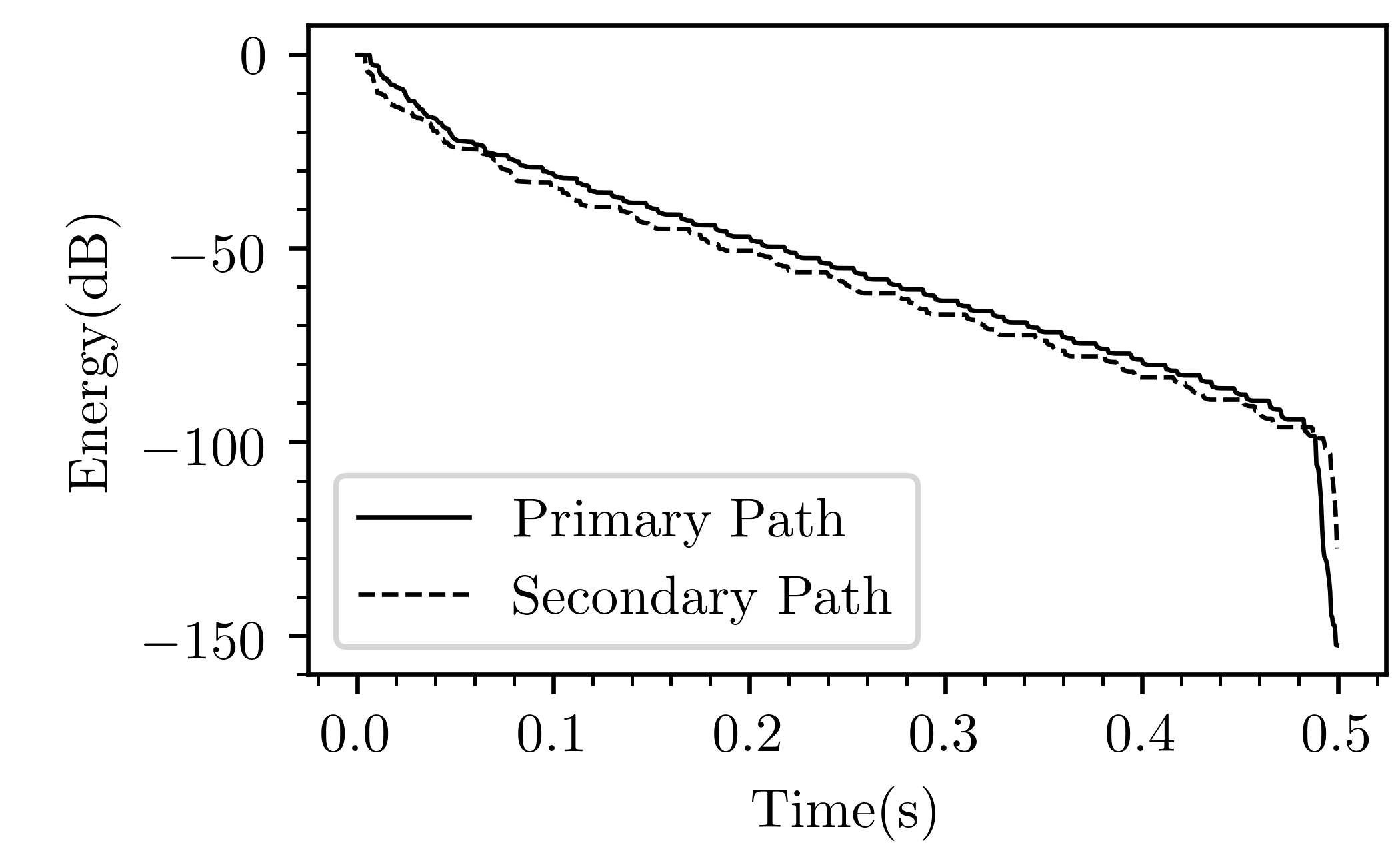}
\caption{Energy Decay Curves of Primary Path $P(z)$ and Secondary Path $S(z)$ for simulation parameters as per Table \ref{tab:SampleCase} }
\label{fig:EnergyDecayCurves}
\end{figure}

Table \ref{tab:SampleCase} shows an arbitrary case for illustration purpose. The noise source is fixed at center of room and the receiving microphone is placed arbitrarily at $(1,3,1.5)$ similar to \cite{barkefors2014design}.  Figure \ref{fig:EnergyDecayCurves} shows energy decay curves of room impulse responses for the arbitrary case.

\begin{table}[!t]
\caption{Simulation settings used for an arbitrary case}
\centering
    \begin{tabular}{|c||c||c|} 
        \hline
            No & Parameter & Value \\
            \hline
                1. & Room size & $6\times4\times3$ meters\\
            \hline
                2. & Noise source location & $(3, 2, 1.5)$\\
            \hline
                3. & Receiver location & $(1, 3, 1.5)$\\
            \hline
                4. & Refection coefficients & $[0.8, 0.7, 0.6, 0.5, 0.4, 0.5]$\\
            \hline
            
       \end{tabular}
    
\label{tab:SampleCase}
\end{table} 

\begin{figure}[!t]
\centering
\includegraphics[width=0.45\textwidth]{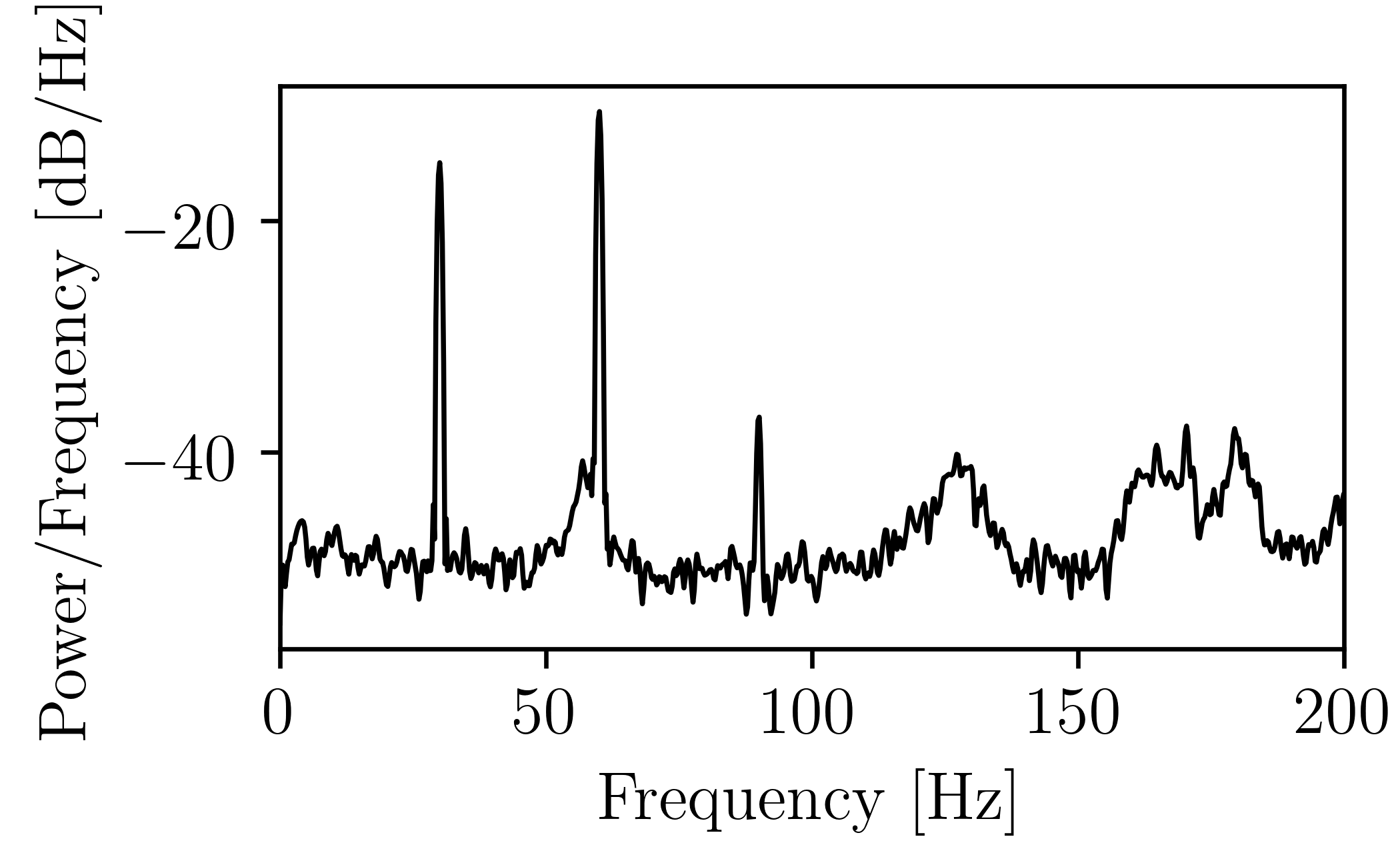}
\caption{Power Spectral Density of source signal used. The narrow band components can be seen at 30 \SI{}{\hertz}, 60 \SI{}{\hertz} and 90 \SI{}{\hertz}. A wide band component of variance 0.1 has been added to the signal.}
\label{fig:SourceSignal}
\end{figure}

The source signal considered is sinusoidal with a wideband component. We used the same method to generate the source signal as used here \cite{bismor2014comments} except the fact that we used open source scientific computing language Python \cite{oliphant2007python} for signal generation and simulation. The method to generate the noise signal including source code can be found in the above reference. Similarly our source codes for the signal generation and simulation is available in the project GitHub repository \cite{ancram}. The sinusoidal components consists of a fundamental frequency of 30 \SI{}{\hertz} and 2 other components which are multiples of the base frequency(i.e, 60 and 90 \SI{}{\hertz}). The signal is generated using Fourier synthesis method. The coefficients of sine components were, -1, -0.5 and 0.1 and that of cosine components were 2, 1 and 0.5. The Power Spectral Density(PSD) of the source signal is shown in Figure \ref{fig:SourceSignal}.

The sampling frequency(Fs) is fixed to \SI{2}{\kilo\hertz} according to Nyquist theorem. The adaptive filter length was fixed as 350 at Fs = \SI{2}{\kilo\hertz}, which is an important parameter determining ANC performance \cite{chang2014feedforward}. A step size of $1\times10^{-5}$ was chosen since we wanted to focus on the level of cancellation achieved rather than how quickly it is achieved. Table \ref{tab:Simulation Settings} shows major settings and parameter values used for the simulation.

\begin{table}[!t]
\caption{Parameter values used for simulation}
\centering
    \begin{tabular}{|c||c||c|} 
        \hline
            No & Parameter & Value \\
            \hline
                1. & Sampling rate & \SI{2}{\kilo\hertz}\\
            \hline
                2. & Room size & $6\times4\times3$ meters\\
            \hline
                3. & Simulation time & \SI{100}{seconds}\\
            \hline
                4. & Sound speed & \SI{343}{meters\per seconds}\\
            \hline
                5. & Reverberation time & \SI{0.4}{seconds}\\
            \hline
                6. & RIR length & 1000 samples\\
            \hline
                7. & Microphone type & omnidirectional\\
            \hline
            
       \end{tabular}

\label{tab:Simulation Settings}
\end{table} 

The assumptions made in this study are as follows. The room is considered to be rectangular in shape with size $6\times4\times3$ meters, without any furniture or moving objects. The reflection coefficient of the six walls are assigned equal and are considered to be frequency independent. The sources and receivers are assumed to be with omnidirectional. The algorithm was simulated for a simulation time of 100 seconds. The RIRs calculated were truncated to 1000 samples corresponding to its reverberation time(T60). The secondary path transfer function which abstracts away reconstruction filter, power amplifier, loudspeaker, the acoustic channel from anti-noise source to receiver, pre-amplifier, anti-aliasing filter and Digital to Analog Converter(DAC) \cite{kuo2006active} is assumed to be known and identified in prior.

\section{Results and Discussion}\label{Results and Discussions}

The resulting signal is analyzed for determining the level of attenuation obtained. The performance measure used for analysis is,  Estimated attenuation($A_{dB}$), similar to \cite{bismor2014comments}, which is calculated using the following formula.

\begin{equation}
A_{dB} = 10\,log_{10}\frac{\mathrm{Var}[d(n)]}{\mathrm{Var}[e(n)]}
\end{equation}

\begin{figure}[!t]
\centering
\includegraphics[width=0.45\textwidth]{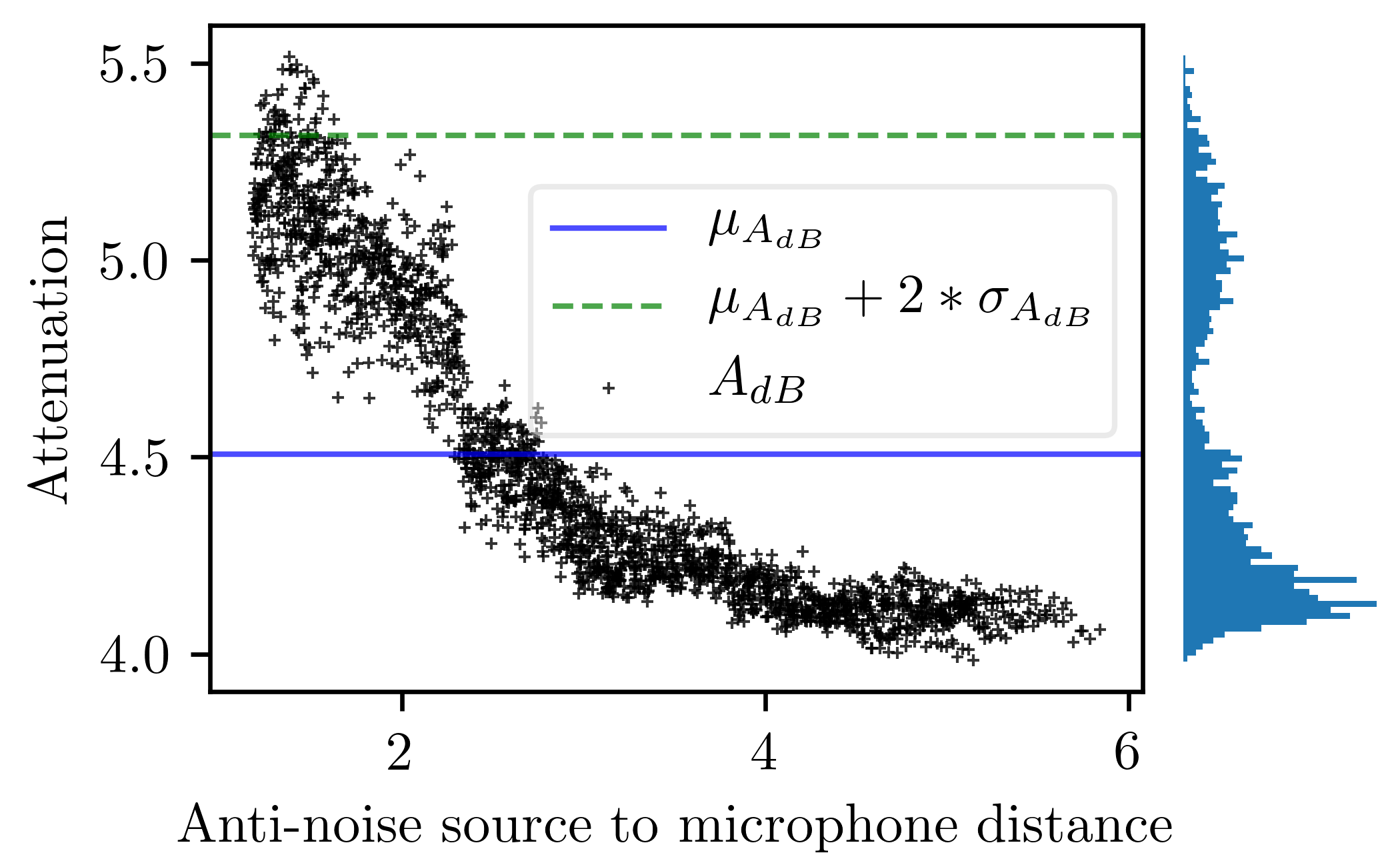}
\caption{Different $A_{dB}$ levels obtained. The simulation settings were as per Table \ref{tab:SampleCase}. It can be seen that the distribution of attenuation levels shown on the right is bi-modal.}
\label{fig:Scatter}
\end{figure}

Where, $\mathrm{Var}[\,.\,]$ represents the mathematical operation of statistical variance. Since our signal of interest is of zero mean, this is equivalent to Mean Square Error(MSE) measure. We consider only the steady state value of error signal, $e(n)$ by trimming off the initial one-third of it. The parameter $A_{dB}$ represents the overall attenuation of the noise signal which is calculated for each of the anti-noise source positions. Figure \ref{fig:Scatter} shows different attenuation levels obtained for different anti-noise source positions as per simulation settings in Table \ref{tab:SampleCase}.

\begin{figure}[!t]
\centering
\includegraphics[width=0.45\textwidth]{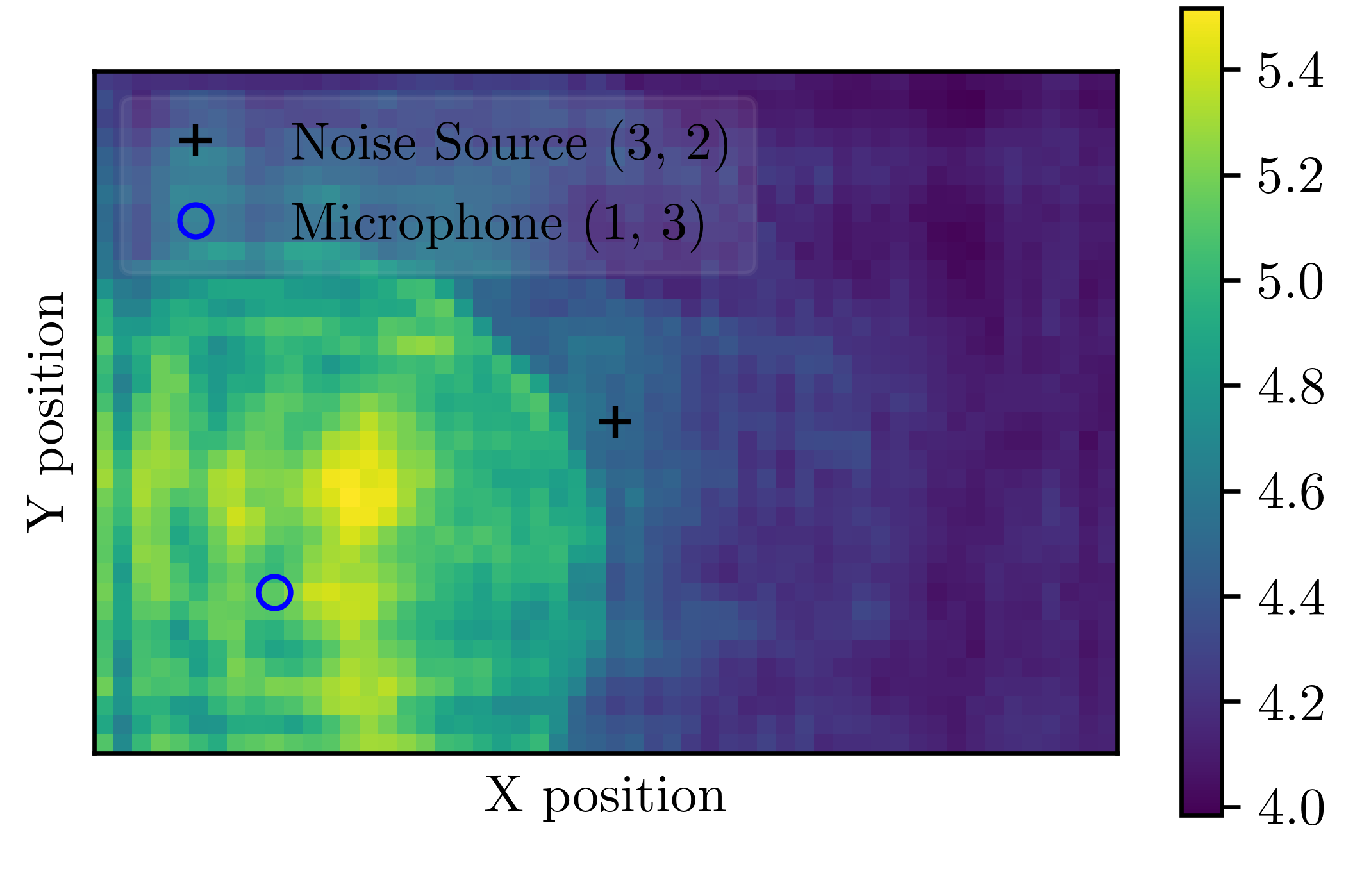}
\caption{Attenuation obtained with respect to each anti-noise source position. A region of locations with good attenuation can be seen around the microphone's location. The simulation settings were as per Table \ref{tab:SampleCase}}
\label{fig:Colorolane}
\end{figure}

The attenuation levels can be visualized as a 2D image, where each point in the image will be representing the steady state attenuation for anti-noise source at that position. Figure \ref{fig:Colorolane} shows the attenuation levels corresponding to different anti-noise source positions as per simulation settings on Table \ref{tab:SampleCase}. 

\begin{figure}[!t]
\centering
\includegraphics[width=0.45\textwidth]{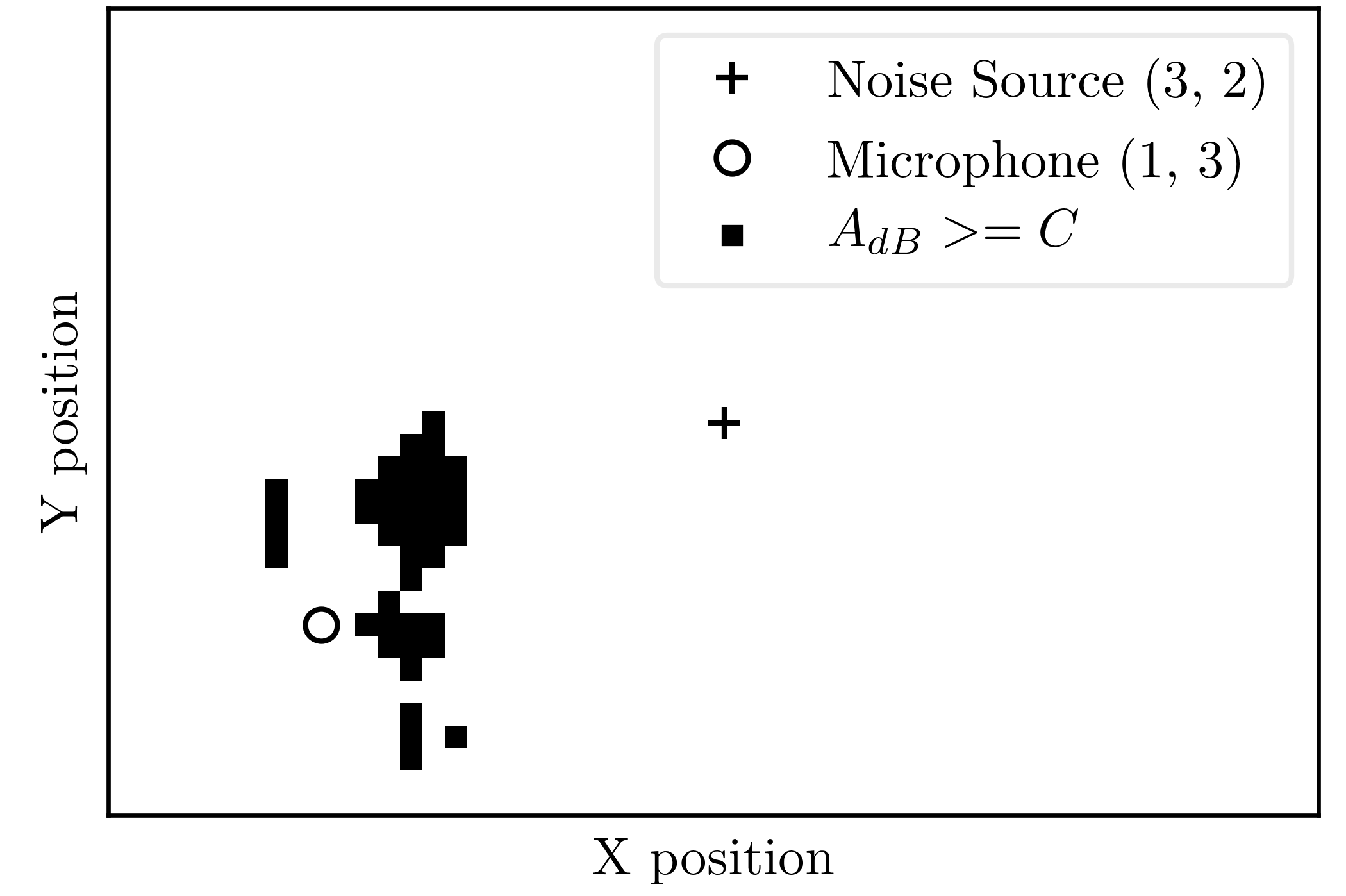}
\caption{Positions representing relatively good attenuation. It can seen that the good anti-noise source positions are located near to error microphone. The simulation parameter values were as per Table \ref{tab:SampleCase}}
\label{fig:ColorolaneSigma}
\end{figure}

To determine good attenuation locations, we chose top $2.5$ percentile attenuation levels by setting a threshold of $C = \mu_{A_{dB}}+1.96\times\sigma_{A_{dB}}$, where $\mu_{A_{dB}}$ and $\sigma_{A_{dB}}$ represents statistical mean and standard deviation of $A_{dB}$ values.  These attenuation values correspond to specific points in the room where the attenuation is relatively high. The corresponding binary figure is shown in the Figure \ref{fig:ColorolaneSigma}. 

\begin{figure}[!t]
\centering
\includegraphics[width=0.45\textwidth]{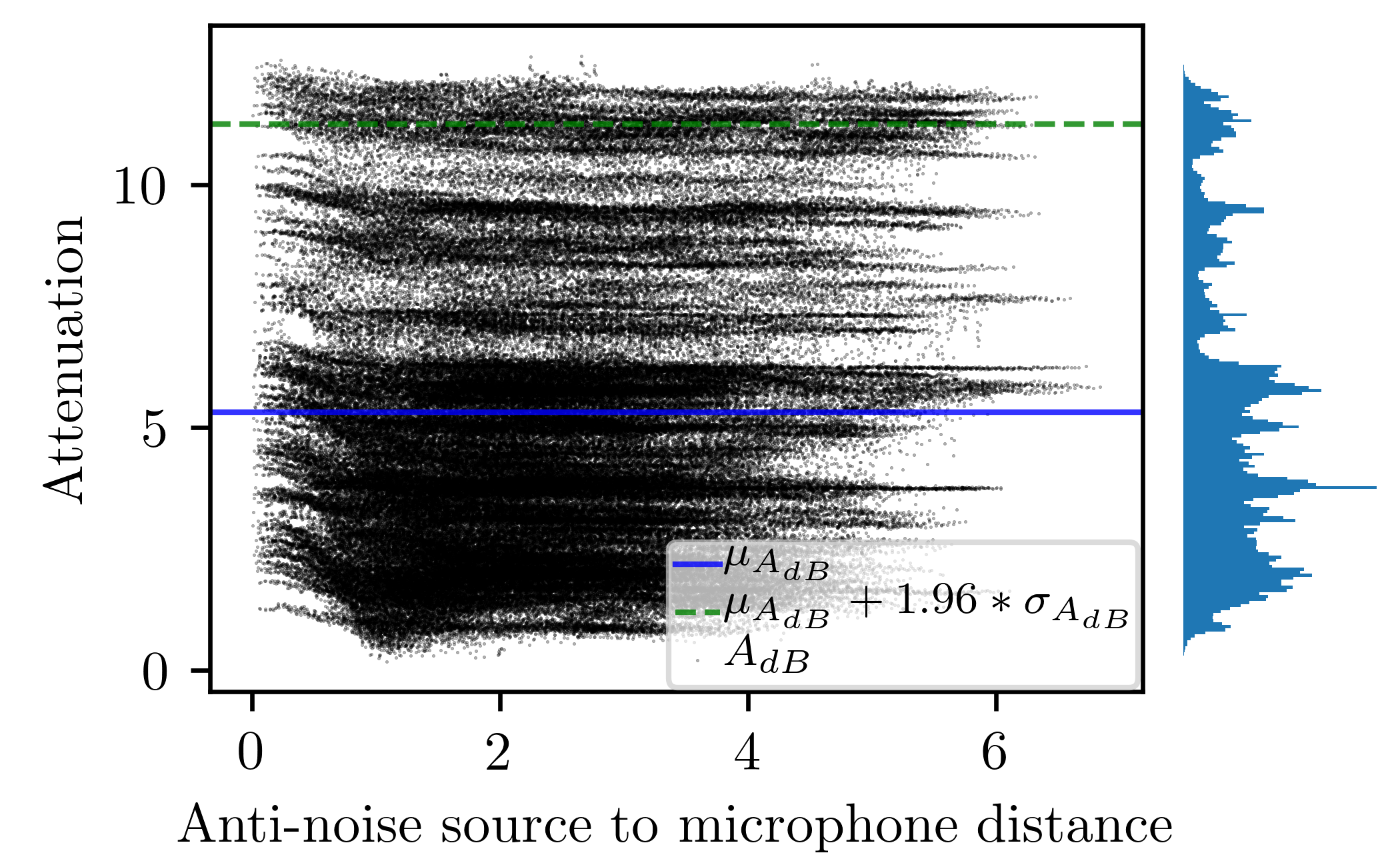}
\caption{Attenuation levels obtained for 100 Monte-Carlo simulations. It can be seen that the average attenuation is 5.31 dB and it can be optimized up to 12.65 dB making a difference of 7.34 dB.}
\label{fig:MasterScatter}
\end{figure}

As can be seen from the Figure \ref{fig:ColorolaneSigma}, the best locations fall near the error microphone location. We think this may due to spectral flatness of secondary path. When the anti-noise source gets closer to the receiver, the secondary path transfer function approaches flat frequency response and the cancellation improves. The attenuation levels of all cases are shown in the Figure \ref{fig:MasterScatter}. It can be seen that the attenuation levels spans in the range [0.17 dB, 12.65 dB] dB, with a mean attenuation of 5.32 dB and 3.02 dB standard deviation.

\begin{figure}[!t]
\centering
\includegraphics[width=0.45\textwidth]{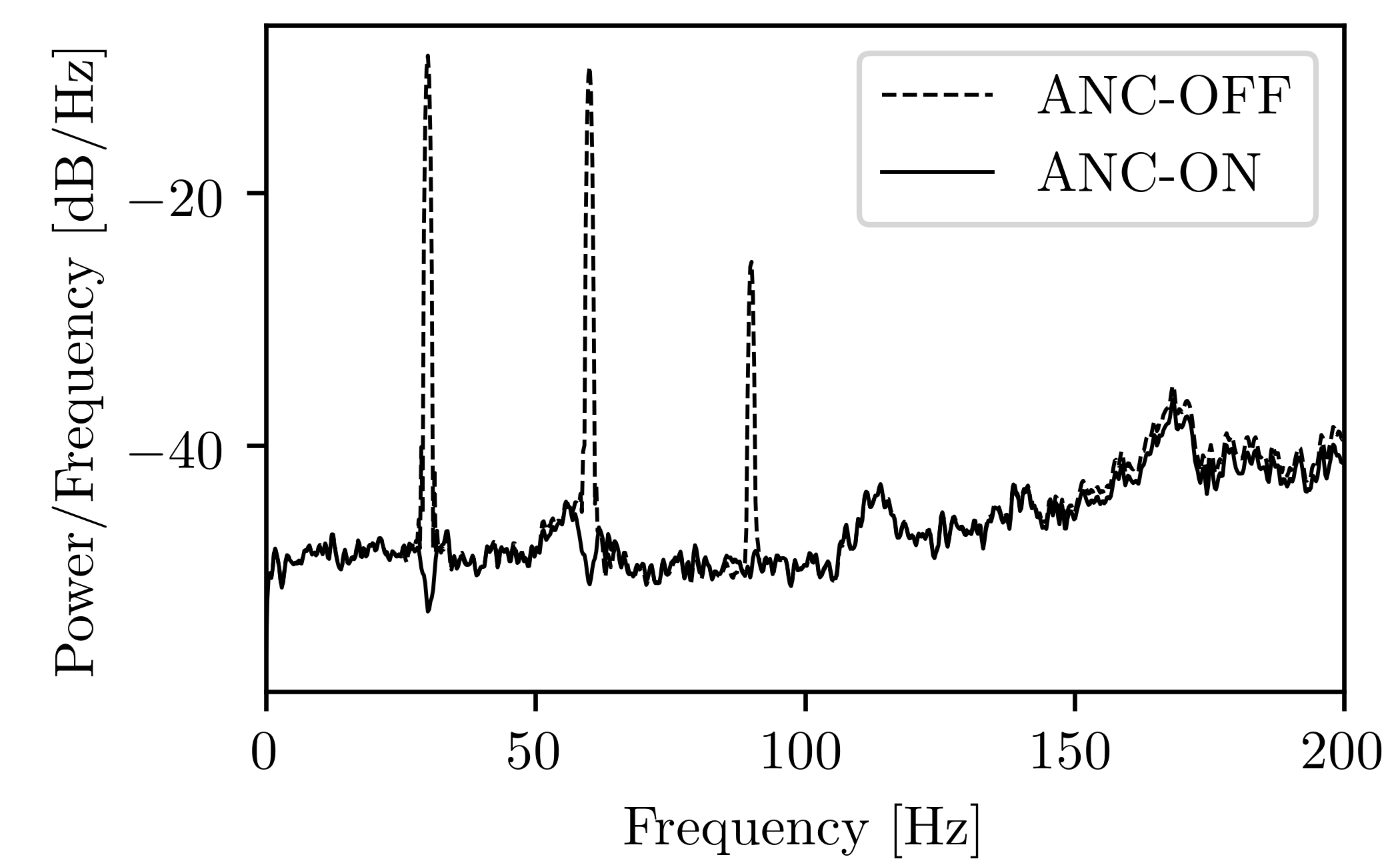}
\caption{Power Spectral Density of $d(n)$ and $e(n)$. It can seen from the graph that the individual narrow band components are significantly attenuated after the the FxLMS algorithm converges to its stable performance.}
\label{fig:FreqDynamics}
\end{figure}

Figure \ref{fig:FreqDynamics} shows the Power Spectral Density(PSD) of the signal before and after cancellation at the optimum anti-noise source position. Our work differs from previous studies in the literature \cite{ANCOptGA} where they concentrate on finding the optimum number of anti-noise sources. Also our results are similar to and reconfirms the findings shown here \cite{kuo2006active}, where they optimize error microphone location by using spectral flatness of secondary path. Compared to these two studies, our method gives insight into best possible geometrical location for anti-noise source using room acoustic models.

The results show the possibility to optimize the anti-noise source location, provided the noise source and receiver positions are known and the RIRs or room reflections coefficients are available. This method has the potential to push further into better cancellation levels for ANC systems. In a realistic scenario, it is possible to calculate reflection coefficients from readily available charts of sound absorption. Also, it is possible to calculate reflection coefficients from RIRs \cite{lehmann2008prediction}. In most cases, room dimensions and shapes are available from blueprints or architecture design diagrams. This opens up the possibility to optimize transducer locations according to any user requirement scenario.

The major drawbacks to our approach are, measurement of real RIRs are time-consuming and tedious. We have not considered time varying changes of the acoustic channels. Also the dynamic variation of the error levels, i.e., how quickly the ANC system attenuates noise have also been not studied. We rather focused on the stable attenuation levels. The secondary path, which is a very important component in ANC systems is assumed to be known and identified. More accurate RIR models will require lengthy RIRs which contributes to computational complexity. There is a tremendous scope for improvement in our work including verifying the results in real lab conditions, improving ease of computation using a more concise representation of RIRs, moving observer or source scenario etc. 

\section{Conclusion}
We have investigated the possibility of using Room Acoustics Models in ANC scenario. We particularly focused on optimizing anti-noise source location once the noise source and receiver are fixed. Results prove that optimizing anti-noise source location improves the level of cancellation by approximately 7.34 dB. This opens up the possibility of optimizing several other physical parameters including, transducer orientation, room shape, etc., in achieving better noise control. Our results show that room acoustic models are good simulation tool which needs more attention by the ANC community.

\section{Acknowledgment}
We are grateful to University Grants Commission(UGC), New Delhi, India for providing necessary funds to do this research, We are grateful to Campus Computing Facility(CCF) and Department of Computational Biology and Bioinformatics, University of Kerala, India for extending the necessary facilities to carry out this research work.


\bibliographystyle{IEEEtran}


\end{document}